# Preconditioned wire array Z-pinches driven by a double pulse current generator

Jian WU [1], Yihan LU [1], Xingwen LI [1], Fengju SUN [2], Xiaofeng JIANG [2], Zhiguo WANG [2],

Daoyuan ZHANG [1], Aici QIU [1,2]

1. State Key Laboratory of Electrical Insulation and Power Equipment, Xi'an Jiaotong University, Shaanxi 710049, China

2. State Key Laboratory of Intense Pulsed Radiation Simulation and Effect, Northwest Institute of Nuclear Technology, Xi'an 710024, China

**Abstract**

Suppressing of the core-corona structures shows a strong potential as a new breakthrough in the X-ray power production of the wire array Z-pinches. In this letter, the demonstration of suppressing the core-corona structures and its ablation using a novel double pulse current generator "Qin-1" facility is presented. The "Qin-1" facility coupled a ~10 kA 20 ns prepulse generator to a ~ 1 MA 170 ns main current generator. Driven by the prepulse current, the two aluminum wire array were mostly heated to gaseous state rather than the core-corona structures, and the implosion of the aluminum vapors driven by the main current showed no ablation, and no trailing mass. The seeds for the MRT instability formed from the inhomogeneous ablation were suppressed, however, the magneto Rayleigh-Taylor instability during the implosion was still significant and further researches on the generation and development of the magneto Rayleigh-Taylor instabilities of this gasified wire array are needed.

Wire array Z-pinch is a powerful and efficient X-ray source of interests for various applications [1,2], such as inertial confinement fusion[3], radiation science[4], warm dense matter[5], and laboratory astrophysics[6]. For an efficient operation of the Z-pinch plasma radiation source, the symmetry and stability of the implosion are desirable to achieve stagnated plasmas with high densities and temperatures. However, the experimental and simulation results reveal that the implosion quality is severely degraded by the development of the magneto Rayleigh-Taylor (MRT) instabilities, and MRT instabilities could be traced back to the wire ablation process of the core-corona structures, a cold dense wire core surrounded by a low density hot corona[7]. The ablation is inhomogeneous but exhibits a strong axial modulation and therefore seeds a large amplitude perturbation at the start of the implosion[8]. The ablation rate of the core-corona structures also affected how the peak power scales with applied current[9].

To suppress the core-corona structures, Lebedev et al proposed the preconditioned wire array using a specially designed load configuration - a two stage wire array [10]. In this configuration, an exploding wire array was employed as a fast current switch to produce a 1 kA, 10 ns prepulse current to preheat the imploding wire array to a gaseous state. After ~ 100ns the main current was applied to the preconditioned wires allowing the implosion of all the mass of the load array. The precursor plasma and the trailing mass were greatly suppressed and the X-ray pulse width was reduced[11]. However, this method by using a two stage wire array was very inconvenience, since it greatly depended on the load configurations. What's more, the time interval between the prepulse current and the main current was limited by the rise time (240 ns) of the current on the MAGPIE facility.

In previous researches, the vaporizations of various metallic wires through electrical explosion were fulfilled [12-14]. Recently, we investigated the possibility of preconditioning of two wires in wire array Z-pinch loads by an auxiliary 1 kA current pulse[15]. In this letter, we report a novel double pulse current generator "Qin-1" facility which couples a ~10 kA 20 ns prepulse generator and a ~ 1 MA 170 ns main current generator. It enables both preheating aluminum wires and imploding the gasified wires with no ablation and no trailing mass, which shows a strong potential as a new breakthrough in the X-ray power production in the wire array Z-pinches.

The prepulse generator used a compact discharge peaking circuit. Its output current to a matched load (5.4 Ω) had a peak value of 10 kA, a rise time of 16 ns and a pulse width of ~ 65 ns, which was sufficient to gasify a wire array composed by ten wires with ~ 10 micrometer diameter. The main current generator was composed by 42 bricks, uniformly distributed in a hexagon shape. Each brick was composed by two 90 nF capacitors (± 100 kV) and one 200 kV gas switch. The outputs of each bricks were connected to a high voltage negative plate, and ferromagnetic cores were not used. The short current of the main current generator reached 800 kA with a 160 ns rising time under a charging voltage of 50 kV. What's more, post-hole structures, posts (prepulse current generator ground) through the negative high voltage plate, were used to realize the prepulse current with a fast rise time and a short pulse width. Since the prepulse current generator and the main current generator were triggered individually, the time delay between the two pulses was adjustable. The typical current waveforms with different time intervals are shown in Figure 1.

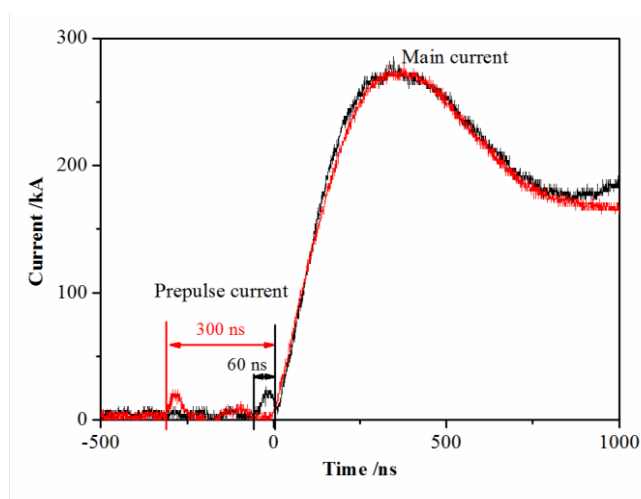

Figure 1 Current waveforms with different time intervals between the prepulse current and the main current under short load. The charging voltage of the main current generator was 30 kV.

Based on the "Qin-1" facility, experiments of the preconditioned aluminum wires were carried out. Two 15μm aluminum wires of 1 cm in length with or without the prepulse current were tested and compared. Rogowskii coils were used to measure the main current and the prepulse current. Optical emissions from the plasma were recorded by a GD-40 phototube. Dynamics of the wires were obtained by the laser shadowgraphy and Mach-Zehnder interferometry[16] using a 150 ps, 532 nm laser beam. The typical laser shadowgraphs of two aluminum wires driven only by the main current from shot 2017042805 (3 mm wire spacing) and shot 2017051304 (5 mm wire spacing) are shown in Figure 2. Quasi-periodic ablation flows and an m=1 unstable precursor plasma were observed in the figure. The inhomogeneous ablations could be considered as one of the important seeds for the MRT instability.

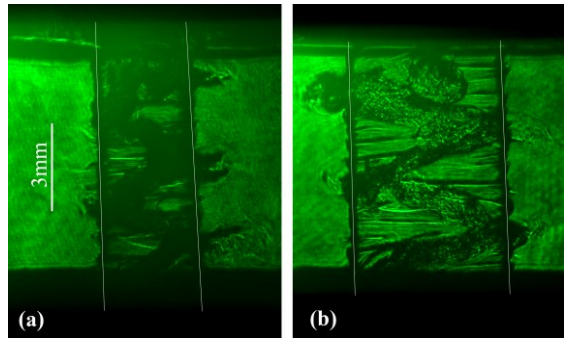

Figure 2 Laser shadowgraphs of two aluminum wires driven by the main current. (a) shot 2017042805, with 3 mm wire spacing and probed at 183 ns after the main current started. (b) shot 2017051304, with 5 mm wire spacing and probed at 160 ns after the main current started.

When the prepulse current was applied, the waveform of the two aluminum wires with 3 mm spacing (shot 2017050901) and its laser probing images are shown in Figure 3 and Figure 4, respectively. The prepulse current had a peak value of 25 kA and a rise time of ~ 20 ns. The main current started at 310 ns later, and its peak value was 250 kA. After the prepulse current was applied, the optical emission from electrical exploding wires could be roughly observed. When the main current started, a much more intense emission was measured.

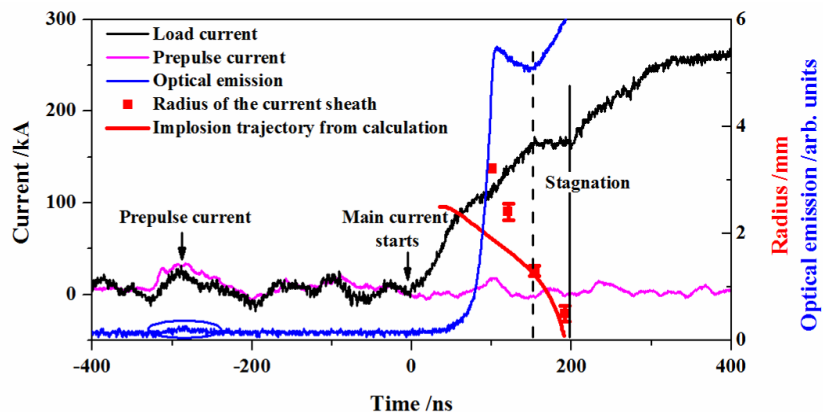

Figure 3   The current and radiation waveform from two aluminum wires preheated by the prepulse current (shot 2017050901). The radius of the current sheath measured from laser shadowgraphs and the imploding trajectory calculated using snowplough model are also presented.

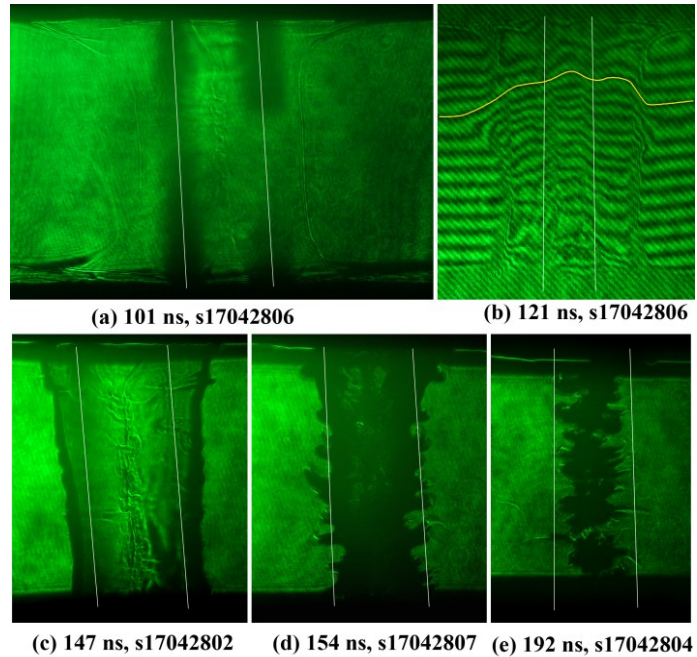

(a) 101 ns, s17042806  (b) 121 ns, s17042806
(c) 147 ns, s17042802  (d) 154 ns, s17042807  (e) 192 ns, s17042804

Figure 4 Laser probing images at different time after the main current started when the prepulse current applied. The wire spacing is 3 mm, and the white lines indicate the initial position of the wires. The probed time and its shot number are labeled below.

The implosion dynamics of the preheated wires was completely different from the ablation process as shown in Figure 2. Figure 4 (a) is the laser shadowgraphs from s17042806, probed in the direction perpendicular to the plane of the two parallel wires, and Figure 4 (b) is the interferometric images obtained at 20 ns later and probed in the direction 30° away from the plane of the two wires. At 101 ns after the main current started, the shadowgram image [Figure 4(a)] shows the exploding products of the two wires in the middle, and a snowplough piston located at 3.2 mm away from the axis. It seems that the magnetic piston at this time only accreted the fast expanding and low density vapors to the axis. These low density vapors formed due to the vaporization and desorbing of the wires by the prepulse current. At 121 ns (20 ns later), the diameter of the magnetic piston decreased to about 2.4 mm, indicating an imploding velocity of the piston of 40 km/s. Based on the shift direction of the fringes from the interferometric images, it was known that the mass within the piston was mainly composed by plasmas. In the middle region of the two wires, the fringes shifts on the axis have a larger value compared to those from the sides, probably indicating that the inner edge of the two exploding wires expanded and collided with each other, resulting in a higher density there. If ignored the neutral atoms, the density distribution of the electrons could be derived from the fringes. The calculation gave an electron areal density on the axis of $10^{18}/cm^2$.

At 147 ns after the current started [Figure 4 (c)], the piston arrived to the position where the main products of the exploding wires located, and small perturbations could be observed at the edges of the current path. Then the piston moved forward to the axis, and MRT instability was significantly increased. At the time around 192 ns after the current started, the plasma stagnated on the axis with a diameter of about 1 mm, and no mass could be observed in the region outside the two wires from the laser shadowgraphs. Based on the position of the piston, the imploding velocity was fit to be about 20 km/s. During this implosion stage, the load current kept unchanged.

Due to the presence of the prefilled mass inside the array, the implosion trajectory of the gasified wires was calculated using the snowplough model. The density distribution of a gasified aluminum wire at the time when the main current started (310 ns after the prepulse current) were calculated using an analytic solution for the expansion of a cylindrical gas column into vacuum[17]. Then the density distribution of the two wires was equivalently averaged over a circle with the total linear mass density and the profile of the density distribution unchanged. Using the averaged mass density distribution and the measured current, the position of the piston varying with time was calculated as presented in Figure 3. In the early time, the radius of the piston was smaller than the data from measurements. The probable reason was that the position of the current sheath in the early time located at the fast expanding and low density vapors, which could not be considered in the model. In the later time (about 147 ns after the current started), current sheath arrived to the position where the main products of the wire located. The imploding trajectory calculated using the snowplough model fitted well with the data from experiments.

In our experiments, it was also found that the implosion dynamics was significantly affected by the wire spacing. As the wire distance increased from 3 mm to 5 mm and with the similar current (prepulse current, main current and their time interval), the typical laser probing images are shown in Figure 5. With the help of the prepulse current, similarly there was no ablation process, and no mass was observed in the region outside the two wires. However, significant differences were also observed in the imploding scenarios. After the main current started, it was observed that both the outer boundary and the inner boundary of each wire had current passed through them [Figure 5(a)]. Then both the inner boundary and the outer boundary imploded to the axis as shown in Figure 5(b). Based on the shadowgraphs at 136 ns and 166 ns, we can estimated the imploding velocity of the inner boundary was no less than 45 km/s, and the velocity for the outer boundary was about 60 km/s. The reason for the different current distribution under a wider spacing was probably that the two gasified wires were still far from each other at the time when the main current applied. Mass inside the wire array was few, and a portion of the current would go along the inner boundarys of the wires.

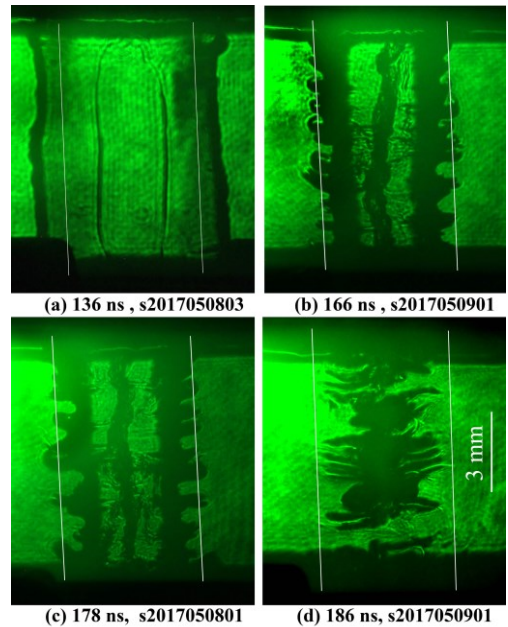

(a) 136 ns , s2017050803  (b) 166 ns , s2017050901
(c) 178 ns, s2017050801  (d) 186 ns, s2017050901

Figure 5　Laser probing images at different time after the main current started when the prepulse current applied. The wire spacing is 5 mm, and the white lines indicate the initial position of the wires. The probed time and its shot number are labeled below.

Significant MRT instabilities were observed only in the outer boundary of the wires. The wavelength of the instabilities was about 1 mm, and the spikes from two wires showed a strong correlation. The spikes also moved inward along with the current sheath, but with a much smaller velocity. Finally the plasma stagnated at about 186 ns after the current started. Although the seeds for the MRT instability formed from the inhomogeneous ablation were suppressed, further researches are needed to investigate the generation and development of the magneto Rayleigh-Taylor instabilities of this gasified wire array.

In summary, we have setup a novel double pulse current generator "Qin-1" facility which couples a ~10 kA 20 ns prepulse generator and a ~ 1 MA 170 ns main current generator, and the experiments on the preconditioned two aluminum wires have been carried out. The two aluminum wires were mostly heated to gaseous state rather than the core-corona structures by the prepulse current, and the implosion of the aluminum vapors showed no ablation, and the trailing mass left behind the initial position of the wires could be neglected. The imploding behaviors of the two gasified wires were affected by their distance at the time when the main current applied. Although the seeds for the MRT instability from the inhomogeneous ablation were suppressed, the MRT instability during the implosion was still significant. Further researches are needed to investigate the generation and development of the magneto Rayleigh-Taylor instabilities of this gasified wire array. In addition, the 1 MA main current generator at present was operated at a level of 200-300 kA restricted by the insulation layer. In the further we would increase the main current by solving insulation problems. Then wire arrays with more wires could be imploded and the effects of the prepulse current on the X-ray power could be compared directly.

This project was supported by the National Natural Science Foundation of China (Grant No. 51237006 and 51407138).